\documentclass[prb,twocolumn,showpacs]{revtex4}
\usepackage{amsmath}
\usepackage{dcolumn}
\usepackage{graphicx}

\begin{document}

\title[Short title for running header]{On the high energy spin excitations in the high temperature superconductors: an RVB theory}
\author{Jiquan Pei Haijun Liao and Tao Li}
\affiliation{Department of Physics, Renmin University of China,
Beijing 100872, P.R.China}
\date{\today}

\begin{abstract}
The high energy spin excitation in the high T$_{c}$ cuprates is
studied in the single mode approximation for the $t-t'-J$ model. An
exact form for the mode dispersion is derived. When the Gutzwiller
projected BCS state is used as the variational ground state, a
spin-wave-like dispersion of about 2.2$J$ is uncovered along the
$\Gamma=(0,0)$ to $\mathrm{M}=(\pi,0)$ line. Both the mode energy
and the integrated intensity of the spin fluctuation spectrum are
found to be almost doping independent in large doping range, which
agrees very well with the observations of recent RIXS measurements.
Together with previous studies on the quasiparticle properties of
the Gutzwiiler projected BCS state, our results indicate that such a
Fermionic RVB theory can provide a consistent description of both
the itinerant and the local aspect of electronic excitations in the
high T$_{c}$ cuprates.
\end{abstract}

\pacs{}

\maketitle

The spin dynamics of the high T$_{c}$ cuprates is a fundamental
issue in the study of high T$_{c}$ superconductivity for two
reasons. On the one hand, the superconducting state of the high
T$_{c}$ cuprates is evolved from and is neighbored by an
antiferromagnetic ordered insulating state. On the other hand, the
d-wave superconduting pairing in the high T$_{c}$ cuprates is widely
believed to be mediated by the antiferromagnetic spin fluctuations.
However, after the intensive studies in the last two decades, it
remains elusive if the spin dynamics of the high T$_{c}$ cuprates is
more appropriately described by an itinerant or a local picture. In
the magnetic ordered parent compounds, both the itinerant and the
local picture work well. However, when the static magnetic order is
melt by doping, it becomes unclear if the local spins remain their
integrity. In the itinerant picture, collective spin fluctuation
emerges only when the system is close to magnetic ordering
instability and is essential only near the magnetic ordering wave
vector. On the other hand, in the local picture, the integrity of
the local spin is dictated by the strong local electron correlation.
It thus survives even when the system is far away from magnetic
instability and when the probing momentum is far away from the
magnetic ordering wave vector. Thus, a detection of the high energy
spin excitations in momentum regions far away from the ordering wave
vector can provide key information on the nature of spin dynamics in
the high T$_{c}$ cuprates.

Most of our understandings on the spin dynamics of the high T$_{c}$
cuprates are obtained from the inelastic neutron scattering(INS)
measurements\cite{Fong,Hayden,Vignolle,Hinkov}. A universal feature
uncovered by these INS studies is the $(\pi,\pi)$ resonance mode in
the superconducting state, which is explained successfully in the
itinerant picture as a spin exciton mode below the superconducting
gap and is taken as a strong support for d-wave pairing. However, as
a result of the limited neutron flux, most INS measurements are made
around the antiferromagnetic ordering wave vector
$\mathrm{Q}=(\pi,\pi)$ and are limited to relatively low excitation
energies.  Very recently, the situation is changed dramatically as
the resonant inelastic X-ray scattering(RIXS) emerges as a powerful
way to study the spin dynamics of the high T$_{c}$
cuprates\cite{Ament}. Together with INS, we are now able to probe
the spin excitation spectrum in the high T$_{c}$ cuprates in a much
larger momentum and energy
range\cite{Tacon11,Vojta,Dean1201,Dean1202,Tacon1301,Dean1301,Dean1302}.
A shocking result from these studies is that the high energy spin
wave excitation of the parent compounds survives even in the heavily
overdoped systems, for which clear signature of electron
itineracy(such as well defined Fermi surface, sharp quasiparticle
peak and well defined d-wave BCS gap) have been confirmed by other
measurements. More specifically, a spin-wave-like dispersive peak is
observed in the spin excitation spectrum by RIXS along the
$\Gamma=(0,0)$ to $\mathrm{M}=(\pi,0)$ line. Apart from a
significant broadening, both the dispersion and the integrated
intensity of this peak are found to be almost doping independent
below $x=0.4$. Here $x$ is the density of the doped holes.

The observation of robust high energy spin wave excitation in the
heavily doped cuprates poses a serious challenge for theoretical
understanding of high temperature superconductivity. From the
strongly correlated point of view, the appearance of dispersive high
energy spin excitation itself is not that surprising, since the
local spins remain their integrity in the doped system. The doped
charge carriers act only to dilute the existing local spins and to
cut short their spatial correlations. At sufficiently short length
scale, we should still expect dispersive spin excitations. However,
it is highly nontrivial how this behavior can be integrated into a
theory which can simultaneously account for the electron itinerancy
around the Fermi surface observed by ARPES
measurements\cite{Damascelli}. For example, in the RPA treatment of
the spin dynamics\cite{Norman}, the spin collective mode is only
well defined around $\mathrm{Q}=(\pi,\pi)$ and is limited to rather
low excitation energies. For momentum far away from
$\mathrm{Q}=(\pi,\pi)$, the spin fluctuation spectrum is essentially
unchanged by the RPA correction and is composed of a broad
particle-hole continuum that can extend in energy up to the band
width(see supplementary material A for an example). Another
important difference between the local and the itinerant picture is
about the local spin sum rule. According to the local picture, the
spin fluctuation spectrum should satisfy the sum rule of $\int
d\mathrm{q} d\omega S(\mathrm{q},\omega)=\frac{3}{4}(1-x)$ as a
result of the no double occupancy constraint. This sum rule is
important for a correct description of the spin dynamics in the
local picture and is violated in the itinerant picture.

The Gutzwiller projected BCS state is generally believed to be a
good variational description of the superconducting state of the
high T$_{c}$ cuprates. After the Gutzwiller projection, the no
double occupancy constraint and the local spin sum rule are enforced
exactly. Previous studies on this state indicate that it offers a
rather good account of the single particle excitation around the
Fermi surface\cite{Gros,Yunoki,Gros06,Nave,Shih,Yang,Li2011}. A
recent projected RPA calculation on this state also explains well
the evolution of the $(\pi,\pi)$ resonance mode with
doping\cite{Li2010}. We thus wonder if the same wave function is
also consistent with the new observations made by recent RIXS
measurements.

In this paper, we calculate the dispersion of the spin excitation in
the single mode approximation for the $t-t'-J$ model with the
Gutzwiller projected BCS state as the variational ground state. The
use of the single mode approximation is just the right choice since
almost all the observed spectral weight in the RIXS measurements are
concentrated in a single broadened peak. We first derive an exact
form of the mode dispersion for the $t-t'-J$ model in the single
mode approximation. We then use the Gutzwiller projected BCS state
to evaluate the expectation values involved in the dispersion
relation. We find when the observed pairing gap is used in the
calculation, a very good agreement with the RIXS observations can be
achieved. This indicates that the Fermionic RVB theory based on the
Gutzwiller projected BCS state can not only account for the low
energy excitation at the energy scale of the pairing gap, but can
also provide an accurate description of the local spin correlation
of the high T$_{c}$ cuprates. It thus provides a balanced account of
the itinerant and the local aspect of the electronic excitations in
the high T$_{c}$ superconductors.

In the single mode approximation, the spin wave excitation at
momentum $\mathrm{q}$ is created by acting on the ground state with
the spin density operator $\mathrm{S}^{+}_{\mathrm{q}}$ and has the
form
\begin{equation}
|\Psi\rangle_{\mathrm{q}}=\mathrm{S}_{\mathrm{q}}^{+}|\mathrm{G}\rangle,
\label{eqn1}
\end{equation}
in which $|\mathrm{G}\rangle$ is the ground state of the system and
is assumed to be normalized. The excitation energy of this mode is
\begin{equation}
\Omega_{\mathrm{q}}=\frac{_{\mathrm{q}}\langle\Psi|H|\Psi\rangle_{\mathrm{q}}}{_{\mathrm{q}}\langle\Psi|\Psi\rangle_{\mathrm{q}}}-\langle
\mathrm{G}|H|\mathrm{G}\rangle,
\label{eqn2}
\end{equation}
in which $H$ is the Hamiltonian of the system. When the spin
fluctuation spectral weight at momentum $\mathrm{q}$ is distributed
in a single broadened peak, as is the case observed in RIXS
measurements along the $\Gamma$-M
line\cite{Tacon11,Dean1201,Dean1202,Tacon1301,Dean1301,Dean1302}, we
can also interpret $\Omega_{\mathrm{q}}$ as the center of gravity of
this broadened peak. The integrated intensity of this peak is given
by the spin structure factor, or
\begin{equation}
I_{\mathrm{q}}=\int d\omega S(\mathrm{q},\omega)=\langle
\mathrm{G}|\mathrm{S}_{-\mathrm{q}}^{-}\mathrm{S}_{\mathrm{q}}^{+}|\mathrm{G}\rangle.
\label{eqn3}
\end{equation}
Here $S(\mathrm{q},\omega)$ is the spin fluctuation spectrum at
momentum $\mathrm{q}$. Thus in the single mode approximation, the
spin fluctuation spectrum is totally determined by the ground state
wave function.

Following well known procedures, it can be shown that the mode
energy $\Omega_{\mathrm{q}}$ can be calculated from the expectation
value of the double commutator between the Hamiltonian and the spin
density operator in the ground state. More specifically, it is given
by
\begin{equation}
\Omega_{\mathrm{q}}=\frac{1}{2}\frac{\langle
\mathrm{G}|[\mathrm{S}_{\mathrm{q}}^{-},[\mathrm{S}_{\mathrm{q}}^{+},H]]|\mathrm{G}\rangle}{\langle
\mathrm{G}|\mathrm{S}_{\mathrm{q}}^{-}\mathrm{S}_{\mathrm{q}}^{+}|\mathrm{G}\rangle}.
\label{eqn4}
\end{equation}
In the derivation, we have assumed that $|\mathrm{G}\rangle$ is the
exact ground state of $H$. Now we apply this formula to the $t-t'-J$
model of the high T$_{c}$ cuprates. The Hamiltonian of the model
reads
\begin{eqnarray}
H=&-&t\sum_{<i,j>,\alpha}(\hat{c}_{i,\alpha}^{\dagger}\hat{c}_{j,\alpha}+\mathrm{h.c.})\nonumber\\
&-&t'\sum_{<<i,j>>,\alpha}(\hat{c}_{i,\alpha}^{\dagger}\hat{c}_{j,\alpha}+\mathrm{h.c.})\nonumber\\
&+&J\sum_{<i,j>}(\vec{\mathrm{S}}_{i}\cdot\vec{\mathrm{S}}_{j}-\frac{1}{4}n_{i}n_{j}),
\label{eqn5}
\end{eqnarray}
in which $\hat{c}_{i,\alpha}=c_{i,\alpha}(1-n_{i,-\alpha})$ is the
constrained Fermion operator that satisfies
$\sum_{\alpha}\hat{c}_{i,\alpha}^{\dagger}\hat{c}_{i,\alpha}\leq1$.
$\alpha=\pm1$ is the spin of the electron. $<i,j>$ and $<<i,j>>$
mean nearest and next-nearest neighboring sites.
$\vec{\mathrm{S}}_{i}=\frac{1}{2}\sum_{\alpha,\beta}\hat{c}_{i,\alpha}^{\dagger}\vec{\sigma}\hat{c}_{i,\beta}$
and
$n_{i}=\sum_{\alpha}\hat{c}_{i,\alpha}^{\dagger}\hat{c}_{i,\alpha}$
are the spin and electron number operators at site $i$. The spin
density operator at momentum $\mathrm{q}$ is given by
$\mathrm{S}_{\mathrm{q}}^{+}=\sum_{i}e^{i\mathrm{q} \cdot
\mathrm{R}_{i}}\mathrm{S}_{i}^{+}$.

To evaluate $\Omega_{\mathrm{q}}$ from Eq.\ref{eqn4}, we need the
commutator between $\mathrm{S}_{\mathrm{q}}^{+}$ and the operators
in $H$. It can be shown directly that these commutators are given
by\cite{Li2010}
\begin{eqnarray}
[\hat{c}_{i,\alpha}^{\dagger},\mathrm{S}_{i}^{+}]&=&-\delta_{\alpha,-1}\hat{c}_{i,+1}^{\dagger}\nonumber\\\relax
[\hat{c}_{i,\alpha},\mathrm{S}_{i}^{+}]&=&-\delta_{\alpha,+1}\hat{c}_{i,-1}\nonumber\\\relax
[n_{i},\mathrm{S}_{i}^{+}]&=&0\nonumber\\\relax
[\mathrm{S}_{i}^{-},\mathrm{S}_{i}^{+}]&=&-2\mathrm{S}_{i}^{z}\nonumber\\\relax
[\mathrm{S}_{i}^{z},\mathrm{S}_{i}^{+}]&=&\mathrm{S}_{i}^{+}.
\label{eqn6}
\end{eqnarray}
When these commutators are inserted into Eq.\ref{eqn4}, we find
\begin{equation}
\Omega_{\mathrm{q}}=\frac{3}{8}\frac{(\mathrm{K}+\frac{16}{3}\mathrm{Ex})(1-\gamma(\mathrm{q}))+\mathrm{K}'(1-\eta(\mathrm{q}))}{S(\mathrm{q})}.
\label{eqn7}
\end{equation}
In the derivation we have assumed that $|\mathrm{G}\rangle$ is spin
rotational invariant. In Eq.\ref{eqn7},
\begin{eqnarray}
\mathrm{K}&=&t\langle\mathrm{G}| \sum_{<i,j>,\alpha}(\hat{c}_{i,\alpha}^{\dagger}\hat{c}_{j,\alpha}+\mathrm{h.c.})|\mathrm{G}\rangle\nonumber\\
\mathrm{K'}&=&t'\langle\mathrm{G}| \sum_{<<i,j>>,\alpha}(\hat{c}_{i,\alpha}^{\dagger}\hat{c}_{j,\alpha}+\mathrm{h.c.})|\mathrm{G}\rangle\nonumber\\
\mathrm{Ex}&=&-J\langle\mathrm{G}|
\sum_{<i,j>}\vec{\mathrm{S}}_{i}\cdot\vec{\mathrm{S}}_{j}|\mathrm{G}\rangle
\label{eqn8}
\end{eqnarray}
are the expectation values of the kinetic energies and the spin
exchange energy in $|\mathrm{G}\rangle$.
$\gamma(\mathrm{q})=(\cos(\mathrm{q}_{x})+\cos(\mathrm{q}_{y}))/2$
and $\eta(\mathrm{q})=\cos(\mathrm{q}_{x})\cos(\mathrm{q}_{y})$ are
the form factors for nearest and next-nearest neighboring hopping
terms in the $t-t'-J$ model.
$S(\mathrm{q})=\sum_{i,j}e^{i\mathrm{q}\cdot(\mathrm{R}_{i}-\mathrm{R}_{j})}\langle\mathrm{G}|\vec{\mathrm{S}}_{i}\cdot\vec{\mathrm{S}}_{j}|\mathrm{G}\rangle$
is the spin structure factor at momentum $\mathrm{q}$.

We thus get a closed form for the mode dispersion for the $t-t'-J$
model in the single mode approximation. From Eq.\ref{eqn7}, it is
clear that the dispersion of the mode is mainly determined by the
momentum dependence of the spin structure factor $S(\mathrm{q})$,
since the expectation values $\mathrm{K}$, $\mathrm{K'}$ and
$\mathrm{Ex}$ in the nominator are all local quantities that are not
sensitive to the exact form of the ground state
$|\mathrm{G}\rangle$. For example, an estimate of these quantities
to the accuracy of 10 percent can be easily achieved by an ordinary
variational wave function. On the other hand, an accurate estimate
of the spin structure factor is much more demanding. The RIXS data
thus provides important information on the spin correlation in the
ground state and may serve as a smoking gun to discriminate between
different theories of the high T$_{c}$ cupartes.

Now we use the Gutzwiller projected BCS state to evaluate the mode
energy $\Omega_{\mathrm{q}}$ and the integrated intensity
$I_{\mathrm{q}}$ for the high T$_{c}$ cuprates. It is given by
\begin{equation}
|\mathrm{G}\rangle=\mathrm{P_{G}}|\mathrm{BCS}\rangle, \label{eqn9}
\end{equation}
in which $\mathrm{P_{G}}$ is the projection into the subspace of no
double occupancy. $|\mathrm{BCS}\rangle$ is the ground state of the
following BCS mean field Hamiltonian
\begin{equation}
H_{\mathrm{MF}}=\sum_{\mathrm{k},\alpha}\epsilon_{\mathrm{k}}c_{\mathrm{k},\alpha}^{\dagger}c_{\mathrm{k},\alpha}
+\sum_{\mathrm{k}}\Delta_{\mathrm{k}}(c_{\mathrm{k},+1}^{\dagger}c_{-\mathrm{k},-1}^{\dagger}+\mathrm{h.c.}),
\label{eqn10}
\end{equation}
in which
$\epsilon_{\mathrm{k}}=-2t_{v}(\cos(\mathrm{k}_{x})+\cos(\mathrm{k}_{y}))-4t'_{v}\cos(\mathrm{k}_{x})\cos(\mathrm{k}_{y})+\mu_{v}$,
$\Delta_{k}=\Delta_{v}(\cos(\mathrm{k}_{x})-\cos(\mathrm{k}_{y}))$.
Here $t'_{v}/t_{v}$, $\mu_{v}/t_{v}$ and $\Delta_{v}/t_{v}$ should
be understood as variational parameters to be determined by energy
optimization. This can be done with the standard variational Monte
Carlo method\cite{Gros}.

\begin{figure}[h!]
\includegraphics[width=7cm,angle=0]{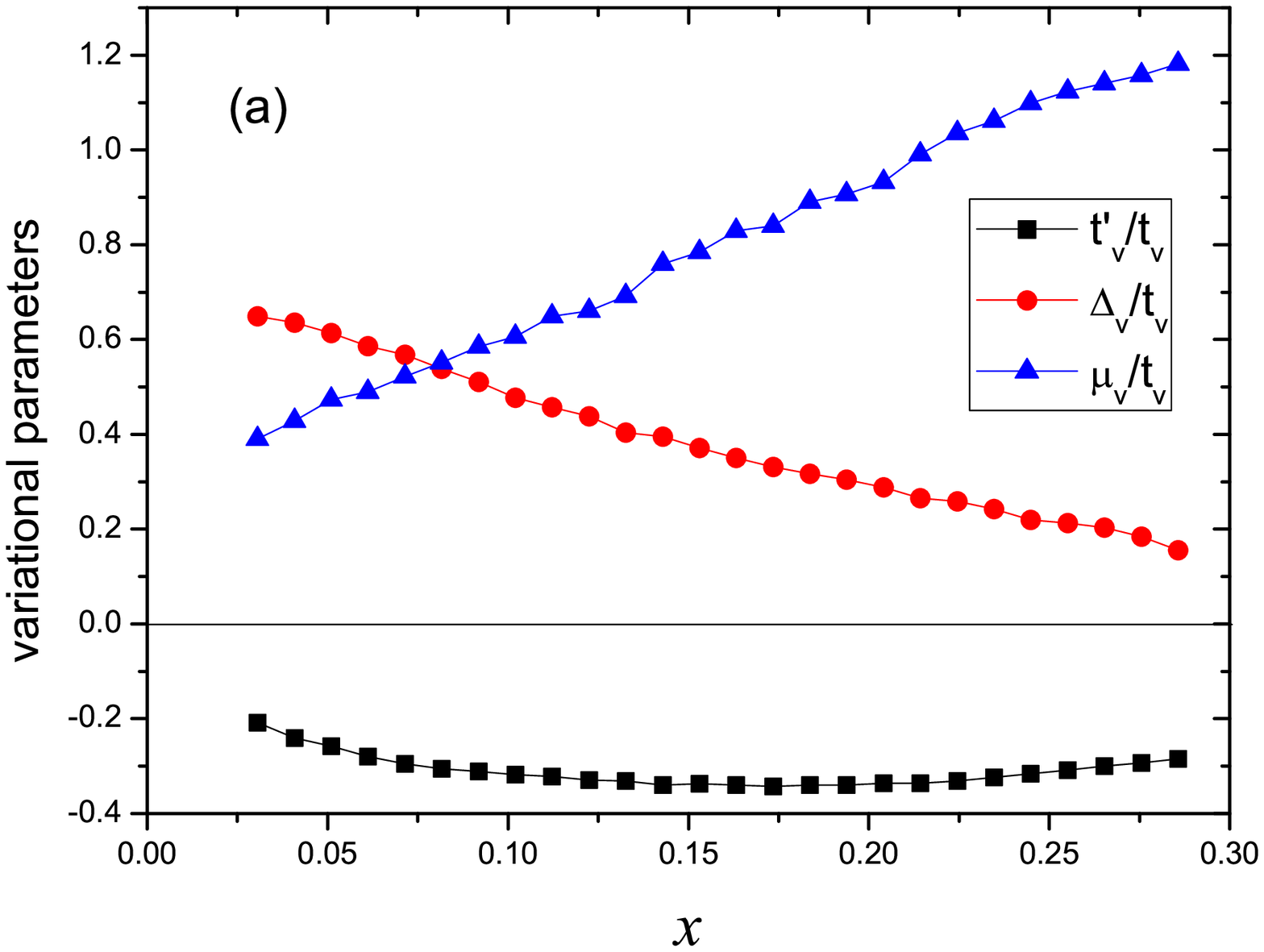}
\includegraphics[width=7cm,angle=0]{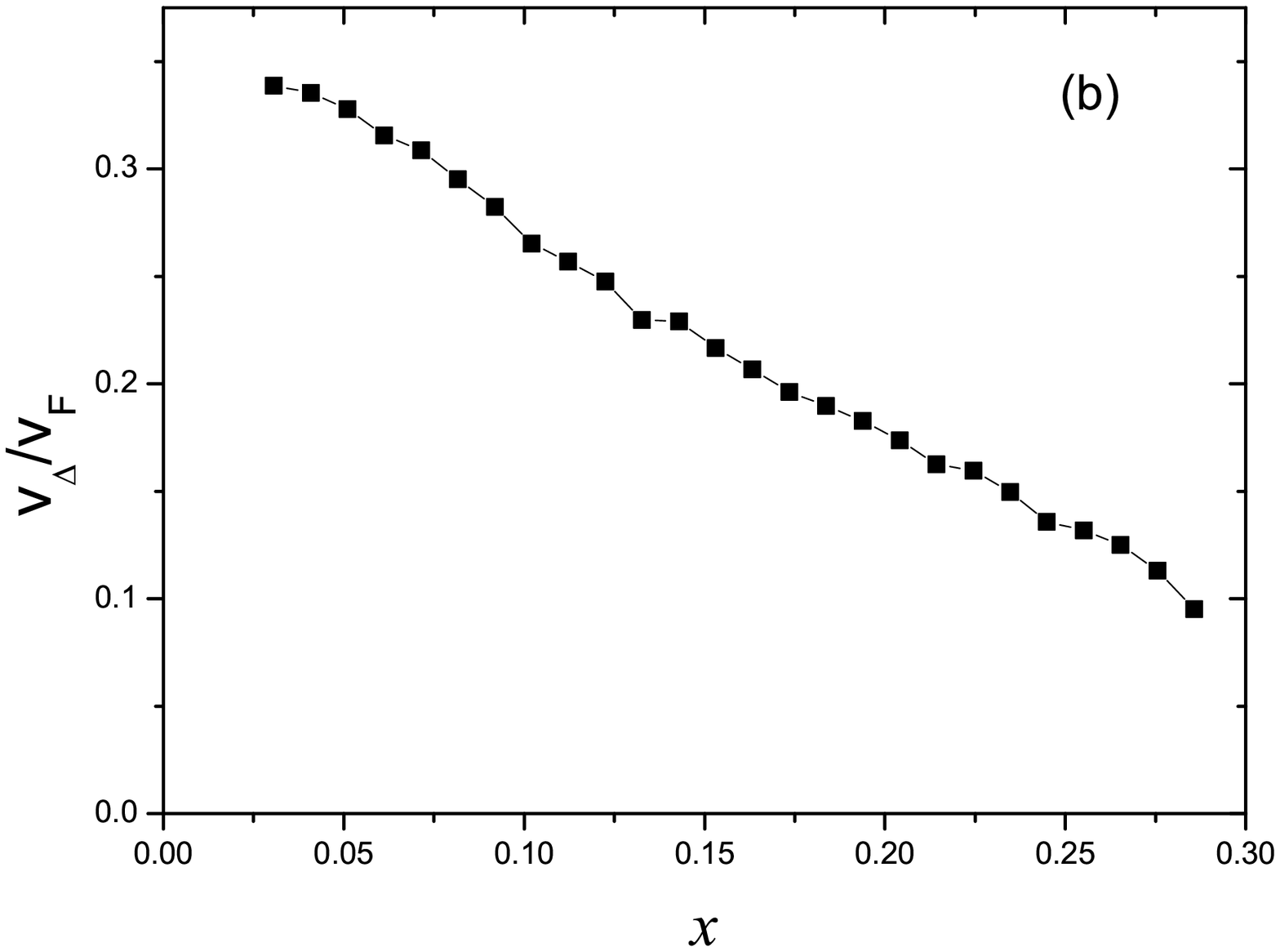}
\caption{(a)The doping dependence of the optimized variational
parameters in the Gutzwiller projected BCS state. (b)The doping
dependence of $\frac{v_{\Delta}}{v_{\mathrm{F}}}$ at the nodal point
as determined from the optimized variational parameters.}
\label{fig1}
\end{figure}

Our calculation is done on a $14\times14$ lattice. We have adopted
the periodic-antiperiodic boundary condition to avoid the d-wave
node. In the calculation, we have set $t'=-0.25t$ and $J=t/3$, as is
usually assumed in the literature. The values of the optimized
variational parameters are shown in Fig.\ref{fig1}a as functions of
doping. For later reference, we also plot in Fig.\ref{fig1}b the
ratio between the gap slope $v_{\Delta}$ and the Fermi velocity
$v_{\mathrm{F}}$ at the nodal point as determined from the optimized
variational parameters. Here the gap slope is defined as
$v_{\Delta}=|\frac{\partial \Delta_{\mathrm{k}}}{\partial
\mathrm{k}}|$. At the nodal point on the Fermi surface, it can be
shown that
$v_{\Delta}=\frac{\Delta_{v}}{2\sqrt{t_{v}^{2}+\mu_{v}t'_{v}}}$.

\begin{figure}[h!]
\includegraphics[width=8cm,angle=0]{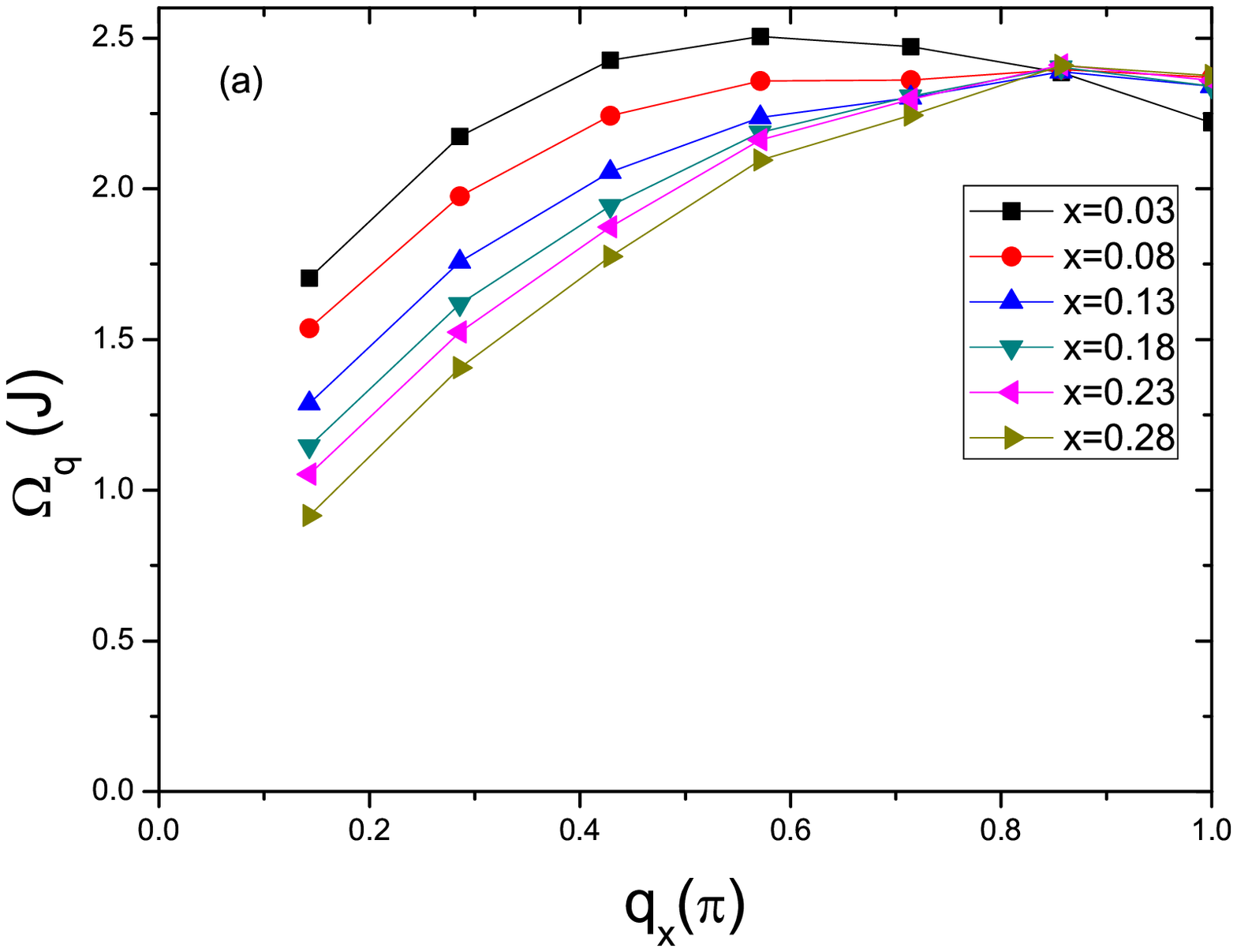}
\includegraphics[width=8cm,angle=0]{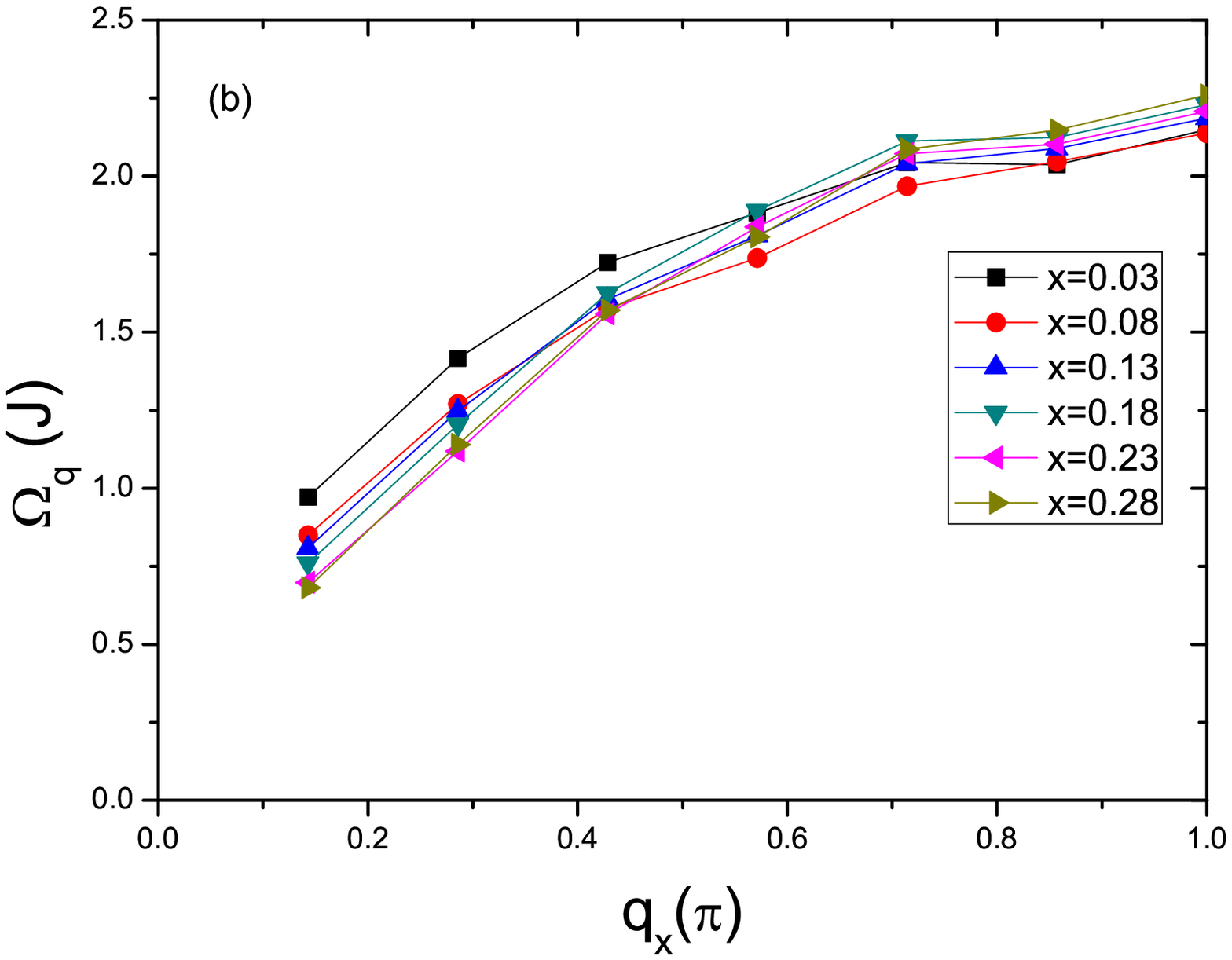}
\caption{The dispersion of the mode energy $\Omega_{\mathrm{q}}$
along the $\Gamma$-M line. Here $\mathrm{q}_{y}$ is fixed at zero.
In (a), we have used the pairing gap determined from variational
optimization. In (b), we have divided the variational pairing gap by
a factor of 4 so as to match the observed value of
$\frac{v_{\Delta}}{v_{\mathrm{F}}}\sim\frac{1}{20}$ around
$x=0.15$\cite{Campuzano,Mesot,Chiao}. } \label{fig2}
\end{figure}

The doping dependence of the mode energy $\Omega_{\mathrm{q}}$ along
the $\Gamma$-M line calculated with the optimized variational
parameters is shown in Fig.\ref{fig2}a. The overall energy scale of
the mode dispersion along the $\Gamma$-M line is about $2J$, which
is quite consistent with the observation in RIXS measurements if we
assume $J=125$ meV\cite{Dean1201}. However, the doping dependence of
the dispersion is rather strong and is inconsistent with the
experimental observations. In particular, the mode energy is found
to increase with decreasing doping at small $\mathrm{q}$ and
deviates strongly from the linear spin wave dispersion around the
$\Gamma$ point at small doping. Such a discrepancy indicates that
certain features in the spin correlation of the high T$_{c}$
cuprates is not correctly captured by the Gutzwiller projected BCS
state if the optimized variational parameters are used.

To see the origin of such a discrepancy, we note the dispersion of
$\Omega_{\mathrm{q}}$ is solely determined by the spin structure
factor $S(\mathrm{q})$. In the long wave length limit, both
$1-\gamma(\mathrm{q})$ and $1-\eta(\mathrm{q})$ in the nominator of
Eq.\ref{eqn7} behave as quadratic functions of momentum. The
coefficients of $1-\gamma(\mathrm{q})$ and $1-\eta(\mathrm{q})$,
namely $\mathrm{K+\frac{16}{3}Ex}$ and $\mathrm{K'}$, are all local
quantities and can both be estimated accurately from the variational
ground state. To reproduce the linear dispersion in the long wave
length limit, the spin structure factor must be linear in
$\mathrm{q}$ in the same limit. This is just what we expect in a
Fermi liquid with a nonzero density of state on the Fermi surface.
More specifically, the spin structure factor in a Fermi gas is given
by
$S(\mathrm{q})=\frac{2}{3}\sum_{\mathrm{k}}\Theta(\epsilon_{\mathrm{k+q}})(1-\Theta(\epsilon_{\mathrm{k}}))$,
in which $\epsilon_{\mathrm{k}}$ is the dispersion of the Fermion.
In the long wavelength limit, only those states that are within a
shell of thickness proportional to $\mathrm{q}$ around the Fermi
surface can contribute to $S(\mathrm{q})$, which is thus
proportional to both $\mathrm{q}$ and the density of state at the
Fermi energy. On the other hand, in the superconducting state, the
low energy spin excitation is gapped out by the superconducting gap
and the spin structure factor is given by
\begin{eqnarray}
S(\mathrm{q})=\frac{1}{2}\sum_{\mathrm{k}}(1-\frac{\epsilon_{\mathrm{k}}\epsilon_{\mathrm{k+q}}+\Delta_{\mathrm{k}}\Delta_{\mathrm{k+q}}}{E_{\mathrm{k}}E_{\mathrm{k+q}}}),
\label{eqn11}
\end{eqnarray}
in which
$E_{\mathrm{k}}=\sqrt{\epsilon^{2}_{\mathrm{k}}+\Delta^{2}_{\mathrm{k}}}$.
In the long wavelength limit, the BCS coherence factor
$(1-\frac{\epsilon_{\mathrm{k}}\epsilon_{\mathrm{k+q}}+\Delta_{\mathrm{k}}\Delta_{\mathrm{k+q}}}{E_{\mathrm{k}}E_{\mathrm{k+q}}})$
is suppressed quadratically as a result of the singlet pairing. Thus
the spin structure factor $S(\mathrm{q})$ is a quadratic function of
momentum for small $\mathrm{q}$. We find the above argument also
apply to the Gutzwiller projected BCS state. This explains why
$\Omega_{\mathrm{q}}$ is gapped at the $\Gamma$ point for small $x$
in Fig.\ref{fig2}a.

With these understandings, we now improve the variational ground
state. For this purpose, we note the pairing gap as determined from
the variational theory is actually much larger than that observed in
experiments. More specifically, both ARPES and transport measurement
find that $\frac{v_{\Delta}}{v_{\mathrm{F}}}\sim\frac{1}{20}$ at the
optimal doping\cite{Campuzano,Mesot,Chiao}, while the value of
$\frac{v_{\Delta}}{v_{\mathrm{F}}}$ determined from the optimized
variational parameters is about 0.2 for $x=0.15$(see
Fig.\ref{fig1}b). Thus, the observed contour of equal energy around
the gap node is about a factor of 4 more anisotropic than that
predicted by the variational theory. For this reason, we scale down
the size of the pairing gap by a factor of four so as to match the
observed value of $\frac{v_{\Delta}}{v_{\mathrm{F}}}$. The mode
energy calculated with the rescaled pairing gap is shown in
Fig.\ref{fig2}b, which is seen to be much more consistent with the
dispersion reported in RIXS measurements. In particular, the doping
dependence of mode dispersion becomes now very weak and an
approximate linear dispersion is recovered in the long wavelength
limit. The overall scale of the dispersion is about 2.2$J$. This
also agrees well with the RIXS data if we assume $J=125$ meV.

In Fig.3, we plot the doping dependence of the integrated mode
intensity $I_{\mathrm{q}}$ along the $\Gamma$-M line. We find
$I_{\mathrm{q}}$ is almost independent of the doping along the whole
$\Gamma$-M line. This is consistent with the observation of RIXS
measurements. However, we note $I_{\mathrm{q}}$ is strongly doping
dependent around the antiferromagnetic ordering wave vector
$\mathrm{Q}=(\pi,\pi)$(see the supplementary material B).

\begin{figure}[h!]
\includegraphics[width=7cm,angle=0]{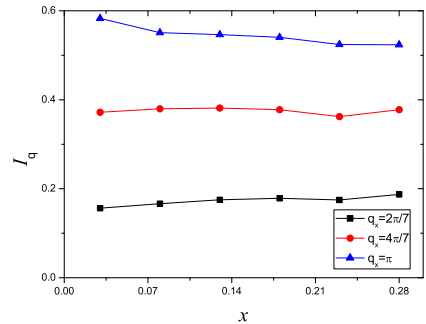}
\caption{The doping dependence of the integrated intensity of the
spin fluctuation spectrum along the $\Gamma$-M line. The parameters
used here are the same as that used in Fig.\ref{fig2}b. }
\label{fig3}
\end{figure}

Thus, the main observations of the RIXS measurements can be well
explained by the Fermionic RVB theory, if the observed pairing gap,
rather than that determined from variational optimization, is
adopted. Together with the results of extensive studies on the
quasiparticle properties of the Gutzwiller projected BCS state, our
results indicate that the Fermionic RVB theory can not only provide
a consistent understanding on the low energy excitations around the
Fermi surface, but can also provide an accurate description of the
local physics of the high-T$_{c}$ cuprates. The Fermionic RVB theory
is thus a promising framework to unify both the itinerant and the
local aspect of electronic excitations in these doped Mott
insulators. Nevertheless, the RIXS data remind us again that the
pairing gap is overestimated substantially in the standard Fermionic
RVB theory. This overestimation is most likely caused by the
insufficient account of the antiferromagnetic spin fluctuation in
the Fermionic RVB theory, which competes with the tendency of spin
singlet pairing. This is reasonable since both the antiferromagnetic
spin fluctuation and the d-wave spin singlet pairing are driven by
the same spin superexchange interaction in the $t-t'-J$ Hamiltonian.
A more advanced theory of the high T$_{c}$ cuprates should of course
treat both ordering tendencies on the same footing.

Finally, we note the single mode approximation becomes less useful
around $\mathrm{Q}=(\pi,\pi)$, as the spin fluctuation spectrum
develops finer structures in this momentum region\cite{Vignolle}(
Supplementary material B contains a more detailed discussion on this
point). A general approach to study these structures in the
Fermionic RVB theory is proposed recently by us\cite{Li2010}. It is
interesting to apply this approach to the high T$_{c}$ cuprates to
arrive at a deeper understanding of such a strongly correlated
system.

This work is supported by NSFC Grant No. 10774187, No. 11034012 and
National Basic Research Program of China No. 2010CB923004.

\section{Supplementary materials}
\subsection{The spin fluctuation spectrum of the high T$_{c}$ cuprates in the itinerant picture}

\begin{figure}[h!]
\includegraphics[width=7cm,angle=0]{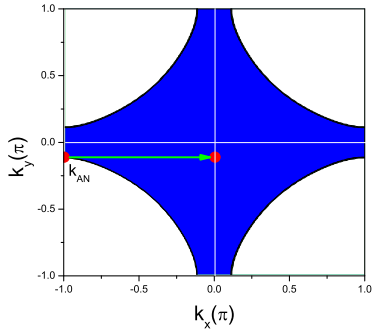}
\caption{The Fermi surface of the tight-binding model we used to
calculate the mean field spin susceptibility
$\chi^{0}(\mathrm{q},\omega)$. The arrow and the red dots illustrate
the transition at the peak energy for $\mathrm{q}=(\pi,0)$. }
\label{fig4}
\end{figure}

To check if the RIXS data is consistent with an itinerant picture,
we have calculated the spin fluctuation spectrum of the cuprates
from the random phase approximation. In the RPA scheme, the dynamic
spin susceptibility is given by
\begin{equation}
\chi(\mathrm{q},\omega)=\frac{\chi^{0}(\mathrm{q},\omega)}{1+J(\mathrm{q})\chi^{0}(\mathrm{q},\omega)},
\label{eqn12}
\end{equation}
in which $\chi^{0}(\mathrm{q},\omega)$ is the bare spin
susceptibility of the band electron and
$J(\mathrm{q})=J(\cos(\mathrm{q}_{x})+\cos(\mathrm{q}_{y}))$ is the
RPA kernel caused by the antiferromagnetic spin exchange. The bare
susceptibility $\chi^{0}(\mathrm{q},\omega)$ in the BCS
superconducting state is given by
\begin{eqnarray}
\chi^{0}(\mathrm{q},\omega)&=&\frac{1}{4}\sum_{\mathrm{k}}(1-\frac{\epsilon_{\mathrm{k}}\epsilon_{\mathrm{k+q}}+\Delta_{\mathrm{k}}\Delta_{\mathrm{k+q}})}{E_{\mathrm{k}}E_{\mathrm{k+q}}})\nonumber\\
&\times&(\frac{1}{\omega-E_{\mathrm{k}}-E_{\mathrm{k+q}}+i0^{+}}\nonumber\\
&-&\frac{1}{\omega+E_{\mathrm{k}}+E_{\mathrm{k+q}}+i0^{+}}),
\label{eqn13}
\end{eqnarray}
in which
$\epsilon_{\mathrm{k}}=-2t(\cos(\mathrm{k}_{x})+\cos(\mathrm{k}_{y}))-4t'\cos(\mathrm{k}_{x})\cos(\mathrm{k}_{y})-\mu$
is the band dispersion and
$\Delta_{\mathrm{k}}=\Delta(\cos(\mathrm{k}_{x})-\cos(\mathrm{k}_{y}))$
is the BCS gap function. The parameters $t$, $t'$ and $\mu$ can be
determined by fitting the ARPES results. Here we set $t=250$ meV,
$t'=-0.3t$, $\mu=-t$. The Fermi surface for this set of parameters
is plotted in Fig.\ref{fig4}, which corresponds to a doping of
$x=0.15$ and is very similar to that observed in the optimally doped
$Bi_{2}Sr_{2}CaCu_{2}O_{8+\delta}$\cite{Damascelli}. We also set
$\Delta=25$ meV, so that
$\frac{v_{\Delta}}{v_{\mathrm{F}}}\approx\frac{1}{20}$. We note that
the conclusion of this section is not sensitive to the exact values
of these parameters.

\begin{figure}[h!]
\includegraphics[width=7cm,angle=0]{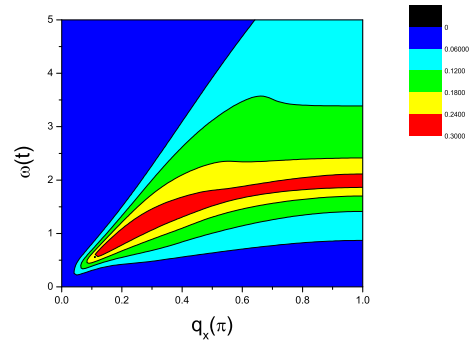}
\caption{A color map of the mean field spin fluctuation spectrum
along the $\Gamma$-M line. Here $\mathrm{q}_{y}$ is fixed at zero. }
\label{fig5}
\end{figure}
\begin{figure}[h!]
\includegraphics[width=7cm,angle=0]{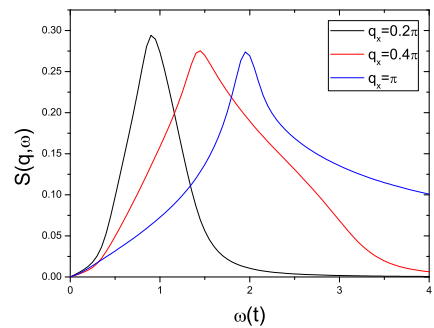}
\caption{The mean field spin fluctuation spectrum at three momentums
along the $\Gamma$-M line. } \label{fig6}
\end{figure}

The bare spin fluctuation spectrum(the imaginary part of
$\chi^{0}(\mathrm{q},\omega)$)is composed of a continuum of the
particle-hole excitations. The RPA correction, which is responsible
for the emergence of soft collective mode around the
antiferromagnetic ordering wave vector $\mathrm{Q}=(\pi,\pi)$, has a
much weaker effect along the $\Gamma$-M line as is probed in the
RIXS measurements. In particular, the RPA correction can not
introduce any new pole in this momentum region. For clarity, we will
only present the mean field spectrum. The result is shown in
Fig.\ref{fig5}. For small $\mathrm{q}_{x}$, the continuum is bounded
above by $v^{m}_{\mathrm{F},x}\mathrm{q}_{x}$, in which
$v^{m}_{\mathrm{F},x}$ is the maximal Fermi velocity in the $x$
direction on the Fermi surface. At the M point, the continuum
becomes as broad as the whole band.

A prominent feature in the mean field spectrum is the dispersive
peak inside the continuum. To understand its origin, we plot in
Fig.\ref{fig6} the spectrum at several momentum along the $\Gamma$-M
line. The peak is found to be related to the Van Hove singularity in
the particle-hole continuum. For $\mathrm{q}$=M, this peak
corresponds to the transition from the antinodal point on the Fermi
surface to momentum around the band bottom. This is illustrated in
Fig.\ref{fig4}. The peak energy at the M point is given by
$E^{peak}_{\mathrm{q}=\mathrm{M}}\approx
E_{\mathrm{k}=\mathrm{k_{AN}}}+E_{\mathrm{k}=0}=|\Delta|+4(t+t')+\mu$.
Here $\mathrm{k_{AN}}$ is the momentum of the antinodal point. For
the parameters we have used,
$E^{peak}_{\mathrm{q}=\mathrm{M}}\approx2t$. Although the dispersion
of this peak looks similar to that observed in RIXS measurements,
the energy of the peak is much higher than that observed.
Furthermore, the peak width is also too broad to be consistent with
experiment. Taken as a whole, the itinerant picture predicts a broad
continuum and a significantly higher energy scale for spin
excitation along the $\Gamma$-M line. The itinerant picture is thus
inconsistent with the RIXS data.

\begin{figure}[h!]
\includegraphics[width=7cm,angle=0]{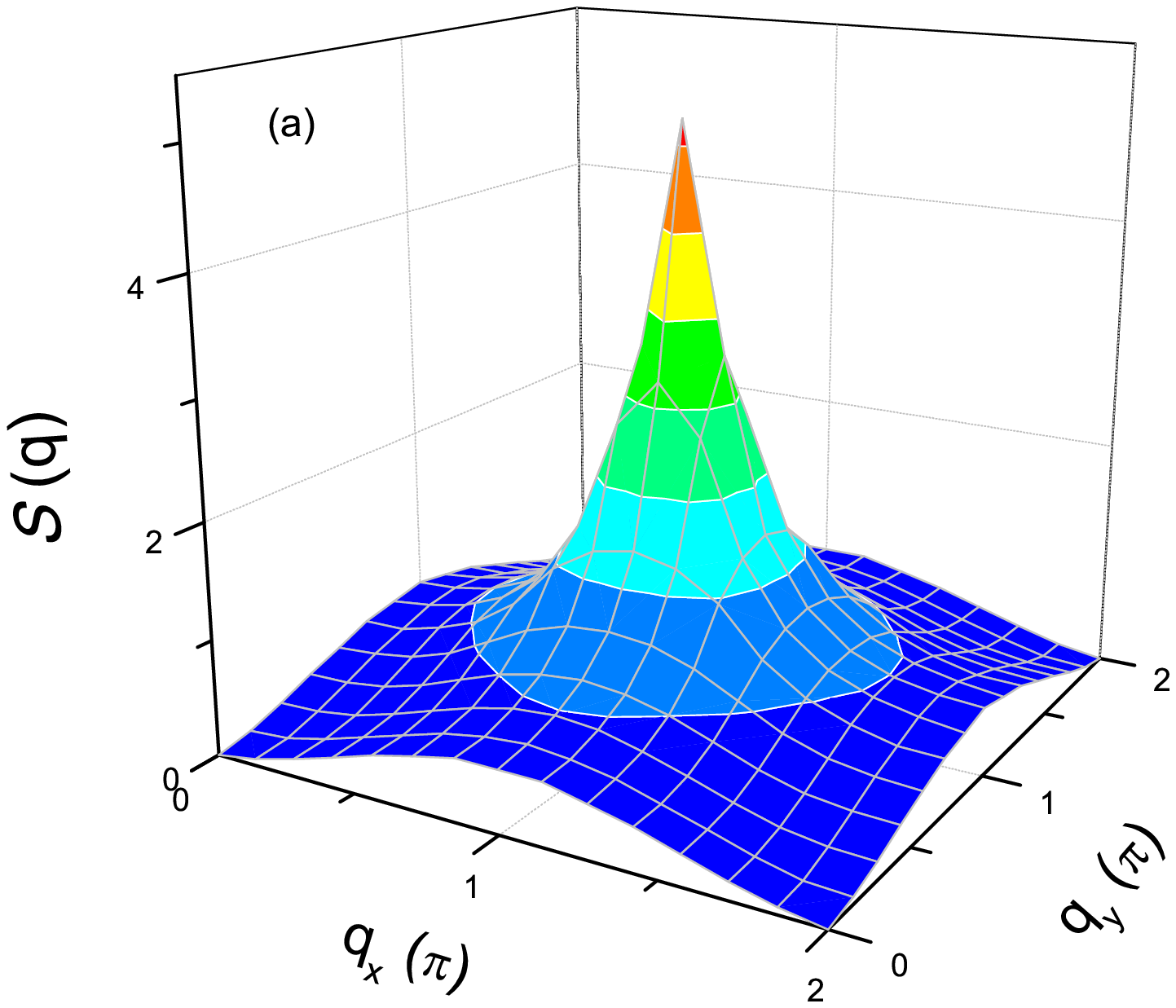}
\includegraphics[width=7cm,angle=0]{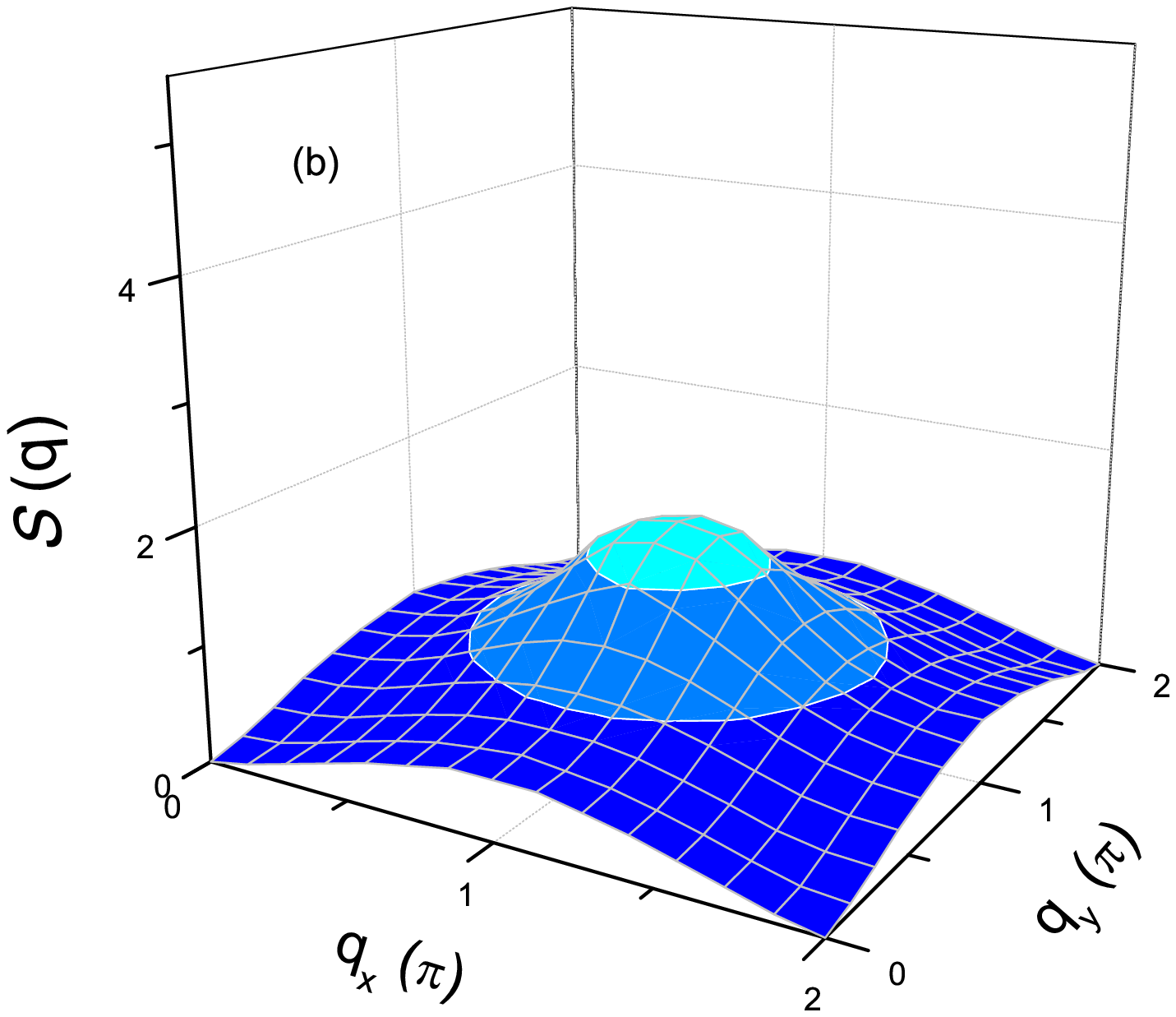}
\includegraphics[width=7cm,angle=0]{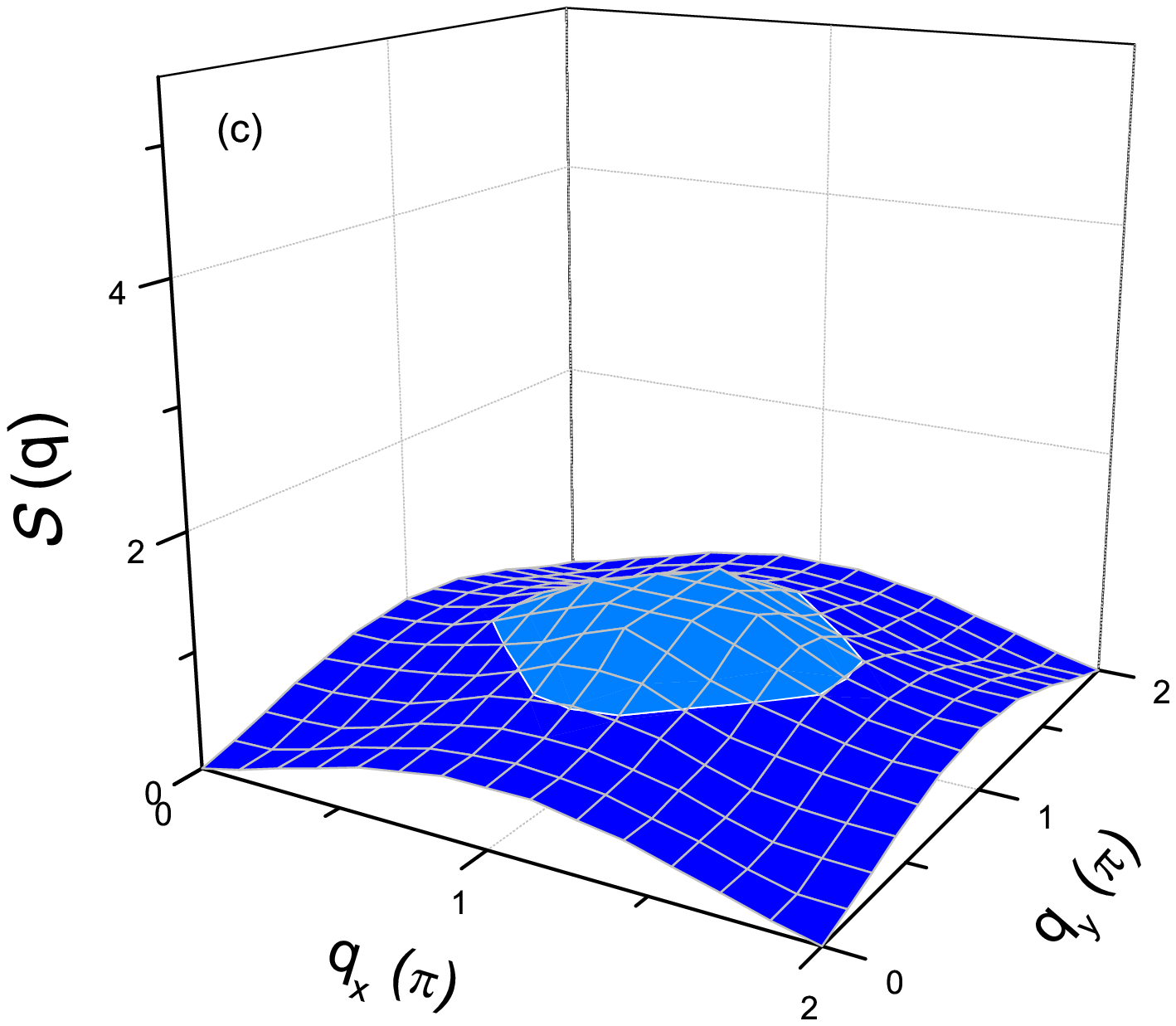}
\caption{The spin structure factor at $x=0.03$(a), $x=0.16$(b) and
$x=0.28$(c). } \label{fig7}
\end{figure}

\subsection{The spin fluctuation spectrum in other region of the Brillouin zone}

\begin{figure}[h!]
\includegraphics[width=7cm,angle=0]{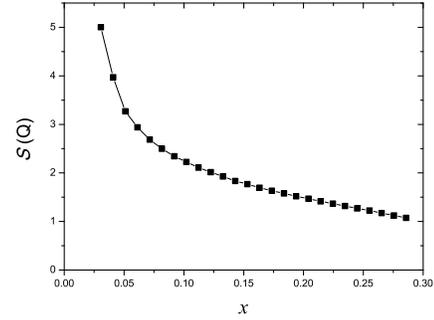}
\caption{The doping dependence of the spin structure factor at
$\mathrm{Q}=(\pi,\pi)$. } \label{fig8}
\end{figure}

From our calculation, we see both the dispersion and the integrated
intensity of the spin excitation spectrum are almost doping
independent along the $\Gamma$-M line. However, this is not
generally true in other regions of the Brillouin zone. In
particular, the spin fluctuation spectrum around the
anfiferromagnetic ordering wave vector $\mathrm{Q}=(\pi,\pi)$ is
strongly doping dependent. In Fig.\ref{fig7}, we plot the spin
structure factor for $x=0.03$, $0.16$ and $0.28$. A strong peak at
$\mathrm{Q}=(\pi,\pi)$ is found for $x=0.03$, which diminishes
rapidly with increasing doping. The doping dependence of the spin
structure factor at $\mathrm{Q}=(\pi,\pi)$ is shown in
Fig.\ref{fig8}.

\begin{figure}[h!]
\includegraphics[width=7cm,angle=0]{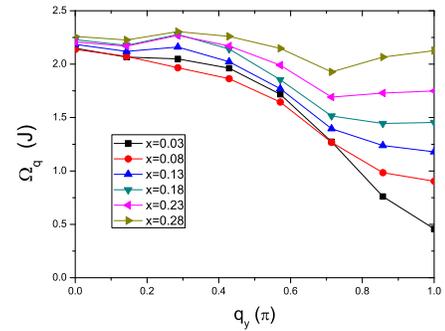}
\caption{The dispersion of $\Omega_{\mathrm{q}}$ along the M-Q line.
Here $\mathrm{q}_{x}$ is fixed at $\pi$. } \label{fig9}
\end{figure}

Apart from the integrated intensity, the center of gravity of the
spin fluctuation spectrum is also strongly doping dependent around
$(\pi,\pi)$. In Fig.\ref{fig9}, we plot the dispersion of
$\Omega_{\mathrm{q}}$ along the the M-Q line. Around
$\mathrm{Q}=(\pi,\pi)$, $\Omega_{\mathrm{q}}$ is found to increase
rapidly with doping and reaches 2$J$ at $x=0.28$. In this region of
the momentum space, the spin fluctuation spectrum develops
multi-component structure\cite{Vignolle} and the single mode
approximation becomes less useful. The large value of
$\Omega_{\mathrm{q}}$ nevertheless indicates that a large amount of
spin fluctuation spectral weight at $\mathrm{q}=(\pi,\pi)$ is
distributed in a broad range of energy of the order of 2$J$ in the
overdoped systems. It would be quite challenging for both INS and
RIXS to detect such strongly smeared-out spectral weight.

\end{document}